# A comment-driven evidence appraisal approach for decision-making when only uncertain evidence available

Shuang Wang, Jian Du

*National Institute of Health Data Science, Peking University*

**Abstract**

**Purpose** In situations like COVID-19 global pandemic, we are in urgent need of evidence to support clinical decision-making, while robust evidence generation and appraisal always lag behind. This study aims to explore whether published research commentaries (such as letters to the editor and editorials) could be used as a quick and reliable evidence appraisal instrument, especially in such situations like COVID-19 that only missing, incomplete, uncertain, or even incorrect evidence is available.

**Methods** Evidence-Comment Networks (ECNs) were constructed for evidence-comment pairs from PubMed indexed COVID-19 publications and the corresponding formal published comments. Top six drugs with the most research publications were included to elaborate the comment-driven evidence assertions, which indicate evidence appraisal results by leveraging the rich information provided by comments towards the primary evidence. Recommendations in WHO guidelines were as gold standard control to validate the accuracy, sensitivity, and efficiency of comments by classifying the sentimental orientations and topics of interest for each comment sentence.

**Main outcome measures** Comment-driven evidence assertions, comment sentimental orientations, comment topics, and the efficiency of comments in shaping knowledge compared with the released recommendations in guidelines.

**Results** For the six drugs included in this study, there are 168 articles as primary evidence being commented and 376 accompanying comments, forming 427 evidence-comment pairs. Comment-driven evidence assertions of 5 drugs were consistent with the recommendations in WHO guidelines when just examining the biggest one or two subgraphs for each drug. The positive/negative sentiments of comments just indicate for/against using a specific intervention in WHO guidelines. Also, comment topics covered all significant points of evidence appraisal, from methodology, clinical themes, to other topics (i.e., ethical issues) and were in accord with the core concerns in suggesting recommendations and even beyond. Furthermore, half of the critical comments emerged 4.25 months on average earlier than guideline release.

**Conclusions** Comment has the potential as a rapid evidence appraisal tool via providing clues to indicate the importance and validity of evidence for decisions. Comment has a selection effect by appraising the benefits, limitations, and other clinical practice issues of concern for existing evidence.

**Introduction**

Decisions are the acts that turn information into action or turn evidence into policy. The need to make effective and efficient health decisions is inescapable, regardless of whether the scientific evidence in place is robust or lacking [1,2]. How to make the best clinical decision from incomplete and uncertain evidence is a common issue, especially in COVID-19 global pandemic[3]. Uncertainty, including probability, ambiguity, and complexity, is always existing in healthcare decision-making scenarios[4]. A summary of data reported that nearly half (47%) of all treatments for clinical prevention or treatment were of unknown effectiveness and an additional 7% involved an uncertain tradeoff between benefits and harms [5].

In addition to the uncertainty about the effects of medical interventions, the timeliness of healthcare decisions needs to be further improved, especially in emergent scenarios. Healthcare decisions primarily rely on clinical practice guidelines and systematic reviews, which synthesize high-quality primary evidence. Nevertheless, as a highly complex process of translating best evidence to best practice, guideline development is time-consuming and requires about 2 to 3 years on average, which is insufficient to satisfy the needs in urgent situations [6]. In the current COVID-19 global pandemic, practitioners must make rapid therapeutic decisions from incomplete, uncertain, and even conflicting scientific evidence. Unproven drugs' use advocated by some physicians and politicians in the early period of COVID-19 may have contributed to the COVID-19 death rates [7]. This devastating global public health crisis reminded scientific researchers and policy makers of the significance of making effective decisions promptly for emergent events even when only limited evidence available.

Most recently, the coevolution of science and policy in COVID-19 has been validated by linking scientific publications and policy documents for both global and China [8,9]. Nevertheless, they failed to reveal the detailed coevolution between scientific evidence and policy recommendations, for example, the selection mechanisms of the included evidence. Science is uncertain inherently. However, policy must be determined. Evidence appraisal, the critical evaluation of published studies, plays an important role in differentiating good science from bad science by uncovering problems in research. To take the first step to propel this complex coevolution, in this study, we proposed to use published research commentaries as a rapid evidence appraisal tool via providing clues to indicate the importance and validity of evidence for therapeutic decisions.

Evidence appraisal is one of the three pivot tasks of a national system to deliver effective healthcare services, along with priority setting and clinical guidelines development [10]. One of the most valuable evidence appraisal approaches is published commentaries, which are expected to play an important role in evidence appraisal [11]. Published research commentaries are formal and short communications that reflect commenters' viewpoints by neutrally commenting on, supporting, or challenging research publications, such as letters to the editor and editorials [12-14]. Two inherent characteristics, diversity (comprehensiveness) and timeliness, made research commentaries crucial in shaping clinical knowledge and indicating details of evidence quality which is imperative in developing clinical guidelines.

First, commentaries evaluate clinical evidence (articles) from diverse perspectives, including the strengths and weaknesses of methodology and clinical aspects. Also, research commentaries may publish original research and case reports that may be insufficient to produce an original research article but cannot be neglected as vital evidence to supplement the field literature [15]. Moreover, commentaries may express audiences' inspiration from original research, which would encourage future studies. These make comments provide a comprehensive and diverse evaluation for a given article or research topic. As an essential part of post-publication peer review, comments were expected to connect current evidence and form the evidence appraisal assertions and even conclusions.

Second, comments were published on time and are much more efficient than the long-time process of developing clinical practice guidelines. For clinical research, the mean and median time from publication of an article to the publication of comment was 6 months and 4 months, respectively. Compared to the slow-developing cycle of clinical guidelines, it responds to clinical circumstances more efficiently and advances the updates of clinical guidelines.

Letters to the editor surrounding the protocols and reported results of randomized controlled trials (RCT), as one type of scientific commentaries, has been analyzed in previous studies. Such type of comment plays post-publication criticism on shaping clinical knowledge [12,16,17]. However, few studies have comprehensively illustrated how exactly research commentaries engage in clinical evidence appraisal and impact the shaping of clinical evidence on a large data scale. That was why this study was initiated. We would like to explore how comments involve in evidence appraisal specifically and to that extent they shape the clinical evidence paths.

Based on prior work, this study was designed to utilize published research commentaries as an evidence appraisal tool in selecting promising research and alerting unsolid studies timely in emergent situations. We used current treatment candidates of coronavirus disease 2019 (COVID-19) for a case study, which is a fantastic example of how comments appraise and shape clinical evidence, and inspire new insights and how it could contribute to the development of clinical guidelines. The present study explores whether comment could be used as a systematic and efficient evidence appraisal assisting tool for evidence-based therapeutic decision-making, especially in such cases that missing, incomplete, uncertain, or even incorrect evidence acquired. Evidence-Comment Networks (ECNs) were constructed for evidence-comment pairs from PubMed indexed COVID-19 publications and the corresponding formal published comments. We aim to identify:

1. Are the sentimental orientations in comment consistent with the recommendation strength in clinical practice guidelines?
2. Are the commentary points aligned with the core concerns of evidence appraisal in developing clinical guidelines (such as the methodological issues of evidence, the clinical adaptability, and other ethical or economic issues)?
3. To what extent are assertions shaped by critical comments faster than the final released recommendations in guidelines?

**Methods**

A workflow diagram is shown in Figure. 1. The whole process of data collection, preprocessing, and networks completion was conducted by Python 3.8.

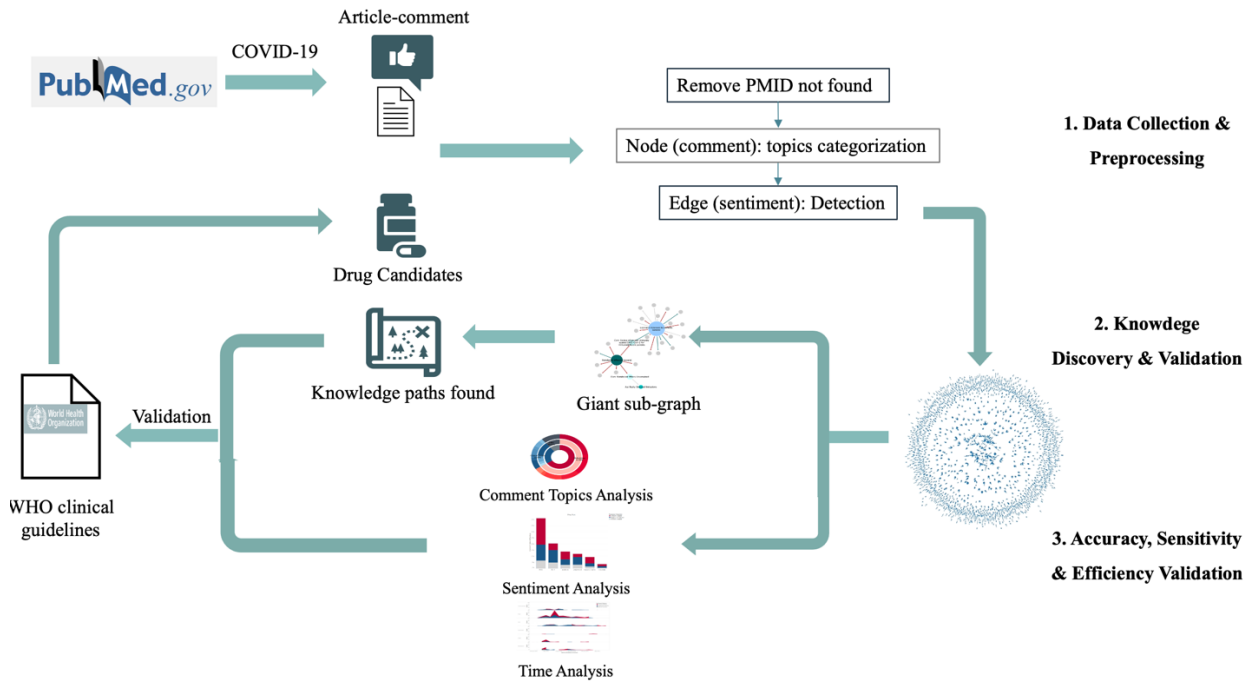

Fig. 1. Workflow Diagram

1. Data Collection

In this study, we utilized PubMed for data analysis and ECNs depiction. To identify COVID-19 related publications and comments, PubMed was queried in two steps on July 21, 2021:

− keyword "(Covid-19[MeSH] OR Covid-19[Title/Abstract]) and hascommentin": identifying publications (evidence) that have COVID-19 in titles or abstracts and include at least one comment (N = 5379)

− keyword "(Covid-19[MeSH] OR Covid-19[Title/Abstract]) and hascommenton": identifying published comments that have COVID-19 in titles or abstracts and comment on at least one publication (N = 5863)

2. Concepts Exploration and Drugs Selection

After COVID-19 evidence-comment pairs dataset was extracted, it is significant to explore which concepts are highly commented. Here we use a text mining tool called PubTator Central, which identified and annotated concepts from the title/abstract sentences of publications as either CellLine, Chemical, DNAMutation, Disease, Gene, Genus, ProteinMutation, SNP, or Species. Concept identifiers come from multiple resources, such as MeSH and gene2pubmed [18]. All concepts with the same MeSH ID were standardized as corresponding MeSH descriptors.

In general, the most common concepts are COVID-19 associated diseases, species, genes (infection mechanism), and drugs. What caught our attention was that three widely discussed drugs during COVID-19 chemicals—hydroxychloroquine, tocilizumab, and remdesivir—were detected from top 5 chemicals-related concepts by PubTator Central. Intuitively, the core section of clinical evidence come from evidence of the clinical treatment process (i.e., drug trials). In addition, as discussed above the unproven drugs usage contributing to COVID-19 deaths and the continuous controversy of the effect of candidate drugs, it would be a great case practice to use drugs as research objects for comment-driven evidence analysis.

As drug has been determined as the research object, we dug deeper for more drug candidates. The top 15 chemicals-related concepts include 9 drugs which were hydroxychloroquine, tocilizumab (IL-6 receptor blockers), remdesivir, vitamin d, azithromycin, chloroquine, lopinavir/ritonavir, steroids, aspirin. To investigate the roles comments played in evidence appraisal, we referred to five versions of "*Therapeutics and COVID-19: living guideline*" by World Health Organization (WHO) and found 5 groups of matched drugs (hydroxychloroquine, IL-6 receptor blockers, remdesivir, lopinavir/ritonavir, and corticosteroid). As Ivermectin was involved in the WHO's series of guidelines, we then added it into research objects list. Finally, six groups of drugs (corticosteroid, remdesivir, hydroxychloroquine, lopinavir/ritonavir, ivermectin, and IL-6 receptor blockers) were chosen as research objects.

3. Data Preprocessing and Labeling

The present work focused on evidence-comment publications of 6 therapeutic drugs for COVID-19 derived from PubMed. First, 17 records were excluded due to 12 PMIDs not being found, resulting in 9296 evidence-comment pairs in total, with 5320 records being commented (4200 primary research publications), and 8483 commentary publications. Next, to focus on six drugs, we extracted evidence-comment data whose titles include any of the six drug names.

Two reviewers (SW and QYG) read all full texts of articles and comments to classify and annotate the topics of interest and the sentimental orientations for each comment. A third reviewer (JD) arbitrated when there was any conflict.

The first is comment topics categorization. To explore the exact commentary points in evidence appraisal, comment topics were categorized based on Kastner et al.'s categorization frame of "Letter to the Editor"[16]. For each comment, one would go through the comment, locate the direct comment sentences, analyze the comment topic categories, and assign the comment to the groups it belongs to. In general, each comment was classified into two comment topic groups hierarchically. The first level categorizations are methodology, clinical themes, and other. Then under each group, comment topics would be further sub-classified, for example, clinical themes - clinical practice related.

Second, once we get to know what those comments were talking about (comment topics), it is of great benefit to figure out the overall comment sentiments, supportive, critical, or neutral. After going through the comments' full texts in the above section, two reviewers would locate the sentences with clear sentiment and label the overall sentimental orientation.

After labeling of all full texts, 21 comments were excluded since some comments were in Spanish and could not be translated accurately or evidence-comment wrongly matched in PubMed. Finally, 168 evidence articles (146 primary research articles; 22 other research articles) and 376 accompanying comments left in this study, made up 427 evidence-comment pairs (354 evidence-comment pairs; 73 comment-comment pairs; 36 repeated records). Evidence-comment pairs had two groups in fact. One group was the real evidence-comment group, representing evidence articles being commented; the other group was the comment-comment group, meaning comments then being commented afterwards. In the followings, we used evidence-comment to represent the all the pairs. Fig. 2 showed the data collection and preprocessing process. The Cytoscape 3.9.0 software was utilized to draw ECNs.

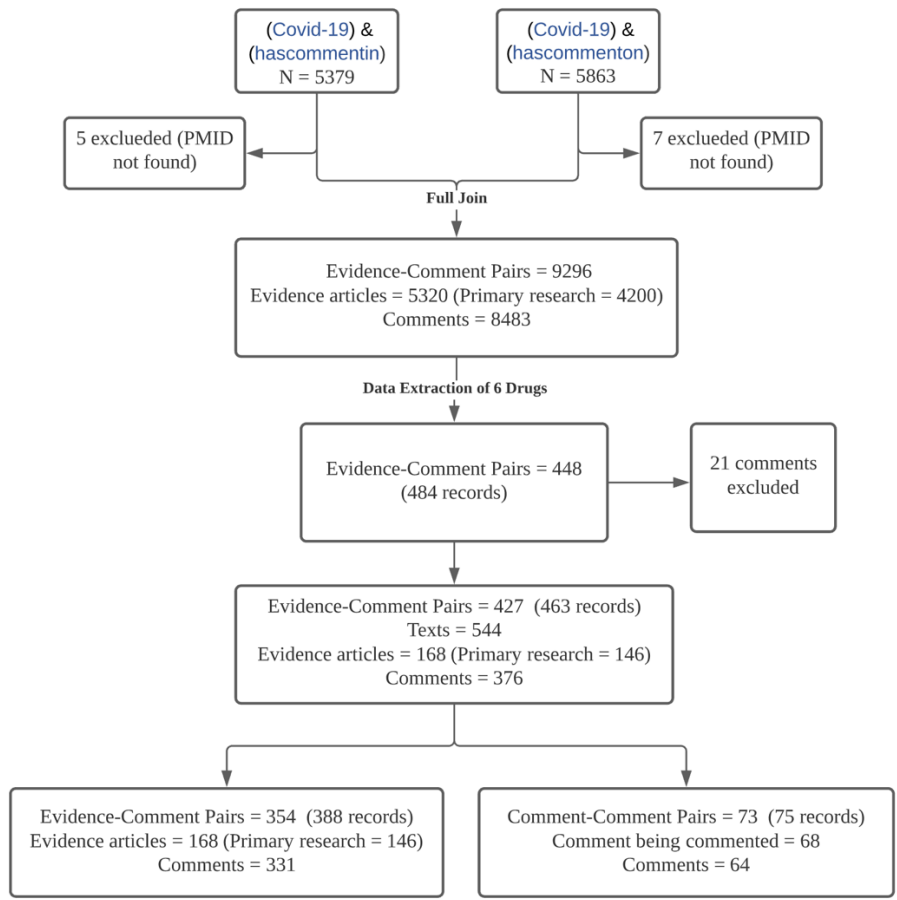

Fig. 2. Data collection and preprocessing

**Results**

1. Descriptive Analysis

We finally included 168 publications relevant to six drugs (146 primary research articles), containing 376 comments, dated from 2003 May to 2021 July, resulting in 427 evidence-comment pairs. The descriptive analysis of 6 drugs were shown in table 1. A series of general characteristics of evidence and comments is summarized, containing evidence timespan, comment timespan, evidence count, comment count, evidence-comment pairs count, the count of subgraphs of each drug, and the count of nodes of the biggest subgraph of each drug.

Table 1. Descriptive analysis of evidence-comment pairs for 6 drugs

| Drugs | Evidence timespan (EPub) | Comment timespan (EPub) | # Evidence | # Comment | Evidence-comment pairs | # Subgraphs | # Nodes (1st subgraph) |
|---|---|---|---|---|---|---|---|
| Corticosteroid | (2020-02-24, 2021-02-18) | (2020-03-20, 2021-06-01) | 17 | 72 | 56 | 19 | 16 |
| Remdesivir | (2020-02-04, 2021-03-18) | (2020-03-04, 2021-04-01) | 19 | 66 | 65 | 16 | 28 |
| HCQ | (2020-02-04, 2021-01-28) | (2020-03-04, 2021-05-27) | 56 | 207 | 190 | 61 | 45 |
| lopinavir/ritonavir | (2003-05-22, 2021-01-07) | (2020-02-24, 2021-03-16) | 12 | 45 | 43 | 11 | 28 |
| Ivermectin | (2020-04-03, 2021-04-03) | (2020-04-16, 2021-07-07) | 4 | 16 | 13 | 7 | 6 |
| IL-6 receptor blockers | (2013-06-01, 2021-03-18) | (2020-03-31, 2021-07-08) | 45 | 98 | 96 | 38 | 18 |

2. WHO guidelines and overall drug evidence-comment networks

2.1. WHO guideline analysis and drugs selection

After matched drugs search, we finally chose the six groups of drugs (corticosteroid, remdesivir, HCQ, lopinavir/ritonavir, ivermectin, and IL-6 receptor blockers) for comment-driven clinical evidence analysis and validated the evidence with WHO guidelines. Till the date of August 12, 2021, WHO has published five versions of COVID-19 therapeutics living guidelines [19]. The recommendation sentences concerning the six drugs are summarized in Fig. 3 below.

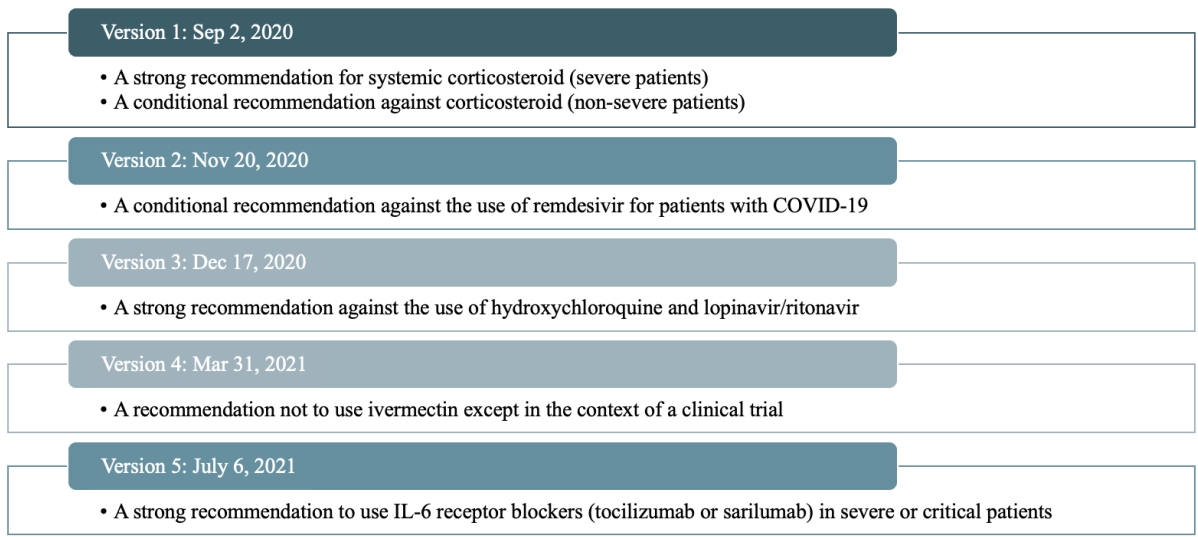

Fig. 3. The evolution of WHO COVID-19 Therapeutics Living Guidelines

2.2. Overall evidence-comment networks

First, overall directed evidence-comment networks of all six groups' drugs were drawn as Fig. 4. PMID was selected as ECNs depiction key features. Each publication's PMID was considered as a node. The relations of evidence-comment pairs were nodes' edges, with arrows pointing from one comment to the article it commented on. A knowledge path was created from one article to the other connected by a bridging comment node.

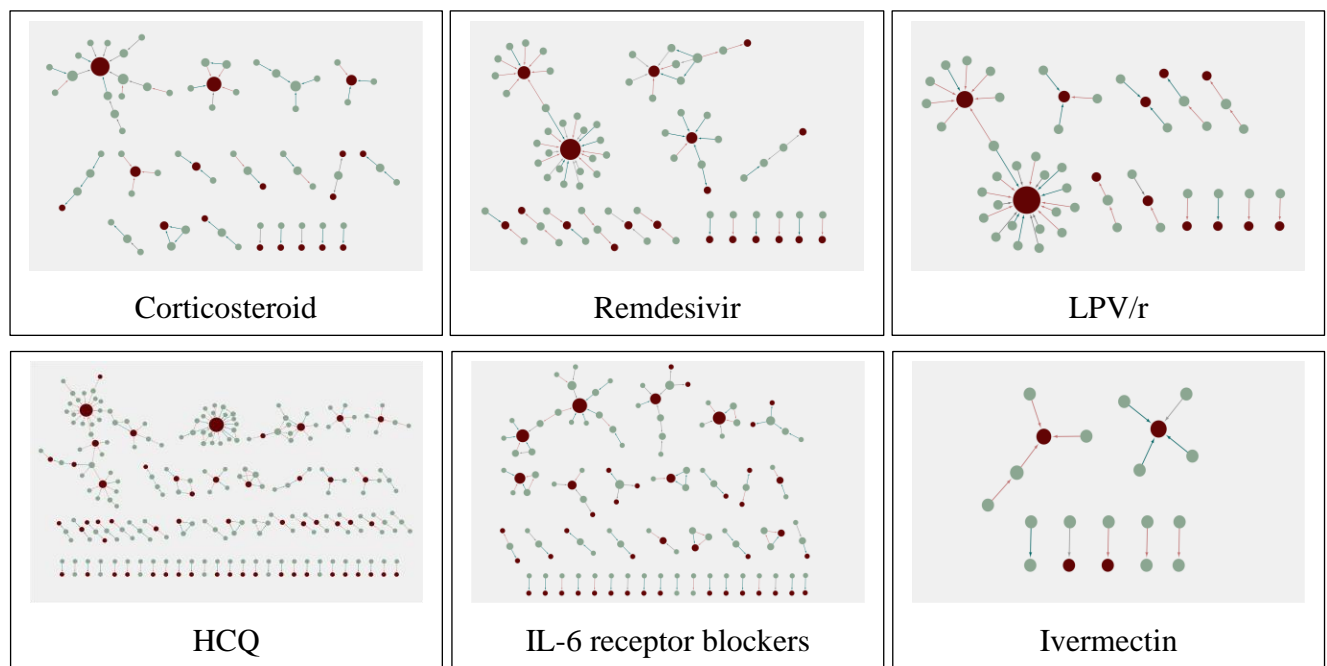

Fig. 4. Overall drug evidence-comment networks | The bigger the node in all connected component networks in each subgraph, the higher the degree centrality is. The red node indicates original research, and the green node indicates non-original research (i.e., review) or comments. Green arrows represent supportive comments, red arrows represent critical comments, and grey arrows represent neutral sentiments.

3. Detailed evidence-comment network analysis

The 1st biggest connected subgraph [20] for each drug was analyzed in-depth separately for elaborating comment-driven evidence assertions since the overall directed ECNs graph was not connected thoroughly. Such assertions in fact indicated the evidence appraisal results by leveraging the rich information provided by comments towards the primary evidence. Comment-driven evidence assertions of each drug's biggest one or two subgraphs was summarized and compared with the final recommendations in WHO guidelines.

3.1. Remdesivir

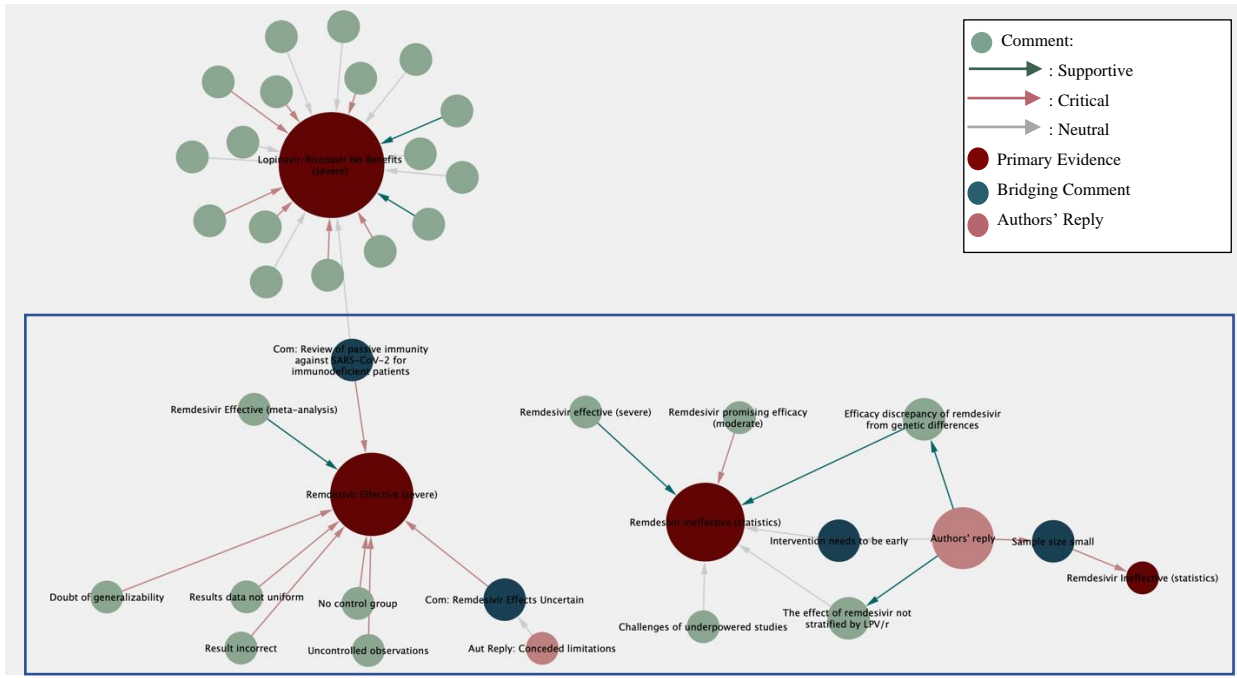

Fig. 5. The effectiveness of remdesivir on COVID-19 was uncertain

The treatment effectiveness of remdesivir remained controversial based on its biggest two subgraphs in Fig. 5. The left subgraph in Fig. 5 is the biggest subgraph of remdesivir. However, it was not about remdesivir itself. It showed a network discussing 1) the effectiveness of remdesivir and 2) the combination effectiveness of lopinavir + ritonavir on COVID-19. Interestingly, the bridging node that connected these two studies is a review-type comment about immunodeficiency. In this comment, Hammarström et al. compared the effects of potential antiviral drugs (i.e., Remdesivir, Lopinavir + Ritonavir) when illustrating immunodeficiency administration in the COVID-19 pandemic. This comparison thus brought these two trials together [21]. Here is Hammarström et al.'s comment sentences:

*"The antiviral drug remdesivir has shown some effects during compassionate use in patients with COVID-19; however, randomized, placebo-controlled clinical trials have yet to prove its value.*

*Another combination of antiviral drugs (lopinavir-ritonavir) did not provide any benefit for hospitalized patients with COVID-19 with severe disease in a randomized, controlled, open-label trial."*

In this evidence-comment network, the comment from Hammarström et al. connected Grein et al.'s position that Remdesivir was effective on severe COVID-19 patients by its compassionate use in 61 patients, to the oppositive position from Cao et al. that no benefits found with lopinavir-ritonavir treatment after a 199 patients' trial [22][23]. This connection introduced both supportive and attacked evidence of the efficacy of antiviral drugs and formed an argument of two sides in this given topic to help understand directly conflicting clinical evidence. Also, Hammarström et al. mentioned that no RCT had proved Grein et al.'s results of the efficacy of remdesivir [21][22].

Further, when we focus on the analysis of remdesivir on the left below, the assertion that remdesivir was effective was undermined and challenged. Grein et al.'s study (Remdesivir, effective) received 7 criticisms out of 8 comments. Among these criticisms, researchers criticized therapy duration as unclear, doubted the generalizability and incorrect higher cumulative incidence of clinical improvement, and the most apparent problem of no control group [24-29]. Grein et al. replied to Wu, Wu, & Lai's "no control group" comment by conceding the limitations of their study and stressed that the results should be interpreted with caution since they were under particular compassionate use.

Nevertheless, it is hard to conclude the effectiveness of remdesivir with this evidence-comment community based on only one scientific evidence. Thus, we further explore the second biggest subgraph of remdesivir on the right side of Fig. 5, which revealed its uncertain effect on COVID-19. Wang et al. conducted an RCT and concluded that remdesivir was not associated with significant clinical improvements in time. Among patients with symptoms in 10 days or less, a numerically faster clinical improvement without statistical significance after receiving remdesivir was detected [30]. This RCT received two supportive, one critical, and three neutral comments. The critical comment refuted that remdesivir was effective for severe patients based on the US National Institutes of Health Adaptive COVID-19 Treatment Trial (ACTT), which concluded that remdesivir could decrease the time to recovery for patients with moderate COVID-19 [31][32].

Based on the leading two subgraphs of remdesivir, more critical comments received than supportive ones undermined the effectiveness of remdesivir, though uncertainties remained. The third version of WHO guidelines on 17 Dec 2020 firstly recommended against using remdesivir and lopinavir on COVID-19, disregarding the severity of diseases. The concrete recommendation here echoed the result of comment-driven evidence assertion [33].

3.2. Lopinavir/Ritonavir

We extracted the top two subgraphs of Lopinavir/Ritonavir (LPV/r) combination to analyze its treatment effect as Fig. 6. As the above section illustrated, the first biggest subgraph of LPV/r was the discussion of the effectiveness of antivirals. The evidence-comment component on the left above was about LPV/r. Cao et al. declared no benefits observed in patients with severe COVID-19 from their RCT, which received more critical comments (38.89, 7/18), and the rest of the comments were either supportive (11.1 %, 2/18) or neutral (55.6%, 10/18) [23]. It seems that comments were more critical about this "no benefits finding."

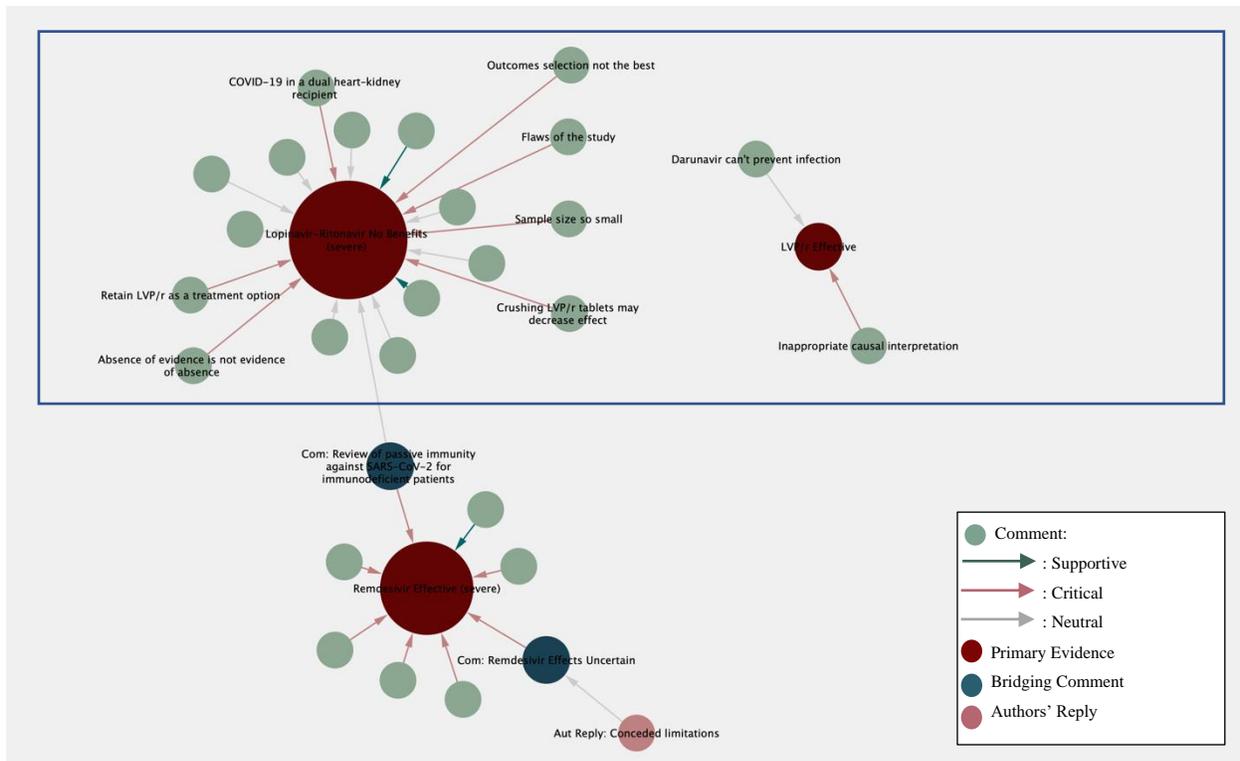

Fig. 6. Treatment efficacy of LPV/r on COVID-19

Negative comments judged a series of methodology errors, such as the "sample size is so small", "crushing LPV/r tablets may weaken effect", "primary endpoint (speed to symptom resolution) may not be the best", and "late therapy and population selection bias" [34-37].

In addition, it is worthy to note that besides pointing out the weakness of this trial, critical comments emphasized that not shutting the door on LPV/r RCT so early since "no benefits observed" does not mean inefficacy and "absence of evidence is not evidence of absence", declaring incremental clinical improvements were crucial even without statistical significance at this urgent time [36-38]. Such calls reflected folks' expectation for any possibility of hope and corresponded to the resume of the WHO SOLIDARITY trial later [39]. In sum, the comment-driven statement from the first subgraph of LPV/r challenged the "no benefits finding" and considered this combination as a promising option though negative evidence was published.

Since only one article was included in this LPV/r research-comment community, we went more profound to the second subgraph on the right for more information. However, evidence to the contrary was found. Lim, et al. [40] reported a case with significantly decreased β-coronavirus viral loads after LPV/r intervention (LPV/r, effective). This case report was criticized by Kim et al. for the inappropriate causal interpretation between laboratory results and therapeutic effects, regarding authors' statement that they have no idea whether the virus decrease is a natural course or antiviral effect [41]. In addition, the other comment weakened this study and expressed that darunavir can not prevent SARS-CoV-2 infection in HIV patients based on their clinical practice related [42]. Lopinavir and darunavir are both HIV protease inhibitors and have similar structures. As a result, the second subgraph showed that the treatment of LPV/r was uncertain or even suspicious.

To sum up, comments of the top 2 biggest subgraphs of LPV/r exhibited contrast attitudes. Given that the first subgraph on the left had more comments than the 2nd subgraph, the integrated comment-driven assertion was that LPV/r was promising though negative evidence existed. Nevertheless, WHO guideline recommended against using HCQ and LPV/r on COVID-19, disregarding the severity of disease [33]. Considering these two graphs cannot represent the overall information of the LPV/r evidence-comment community, new evidence clues may reverse the assertion as keeping digging. This turned out the uncertainty of science exploration. Whereas, as we were exploring deeper, we were getting closer.

3.3. The combination of hydroxychloroquine & azithromycin

A relevant absolute conclusion was not reached from the above evidence-comment networks. However, a more conclusive claim could be developed as more evidence-comment pairs joined together with consistent sentiments. In Fig. 7, the efficacy of hydroxychloroquine & azithromycin (HCQ + A) on COVID-19 is critically negated. Among these comments, one pointed out relevant research evidence (hydroxychloroquine & azithromycin, ineffective) to prove own view.

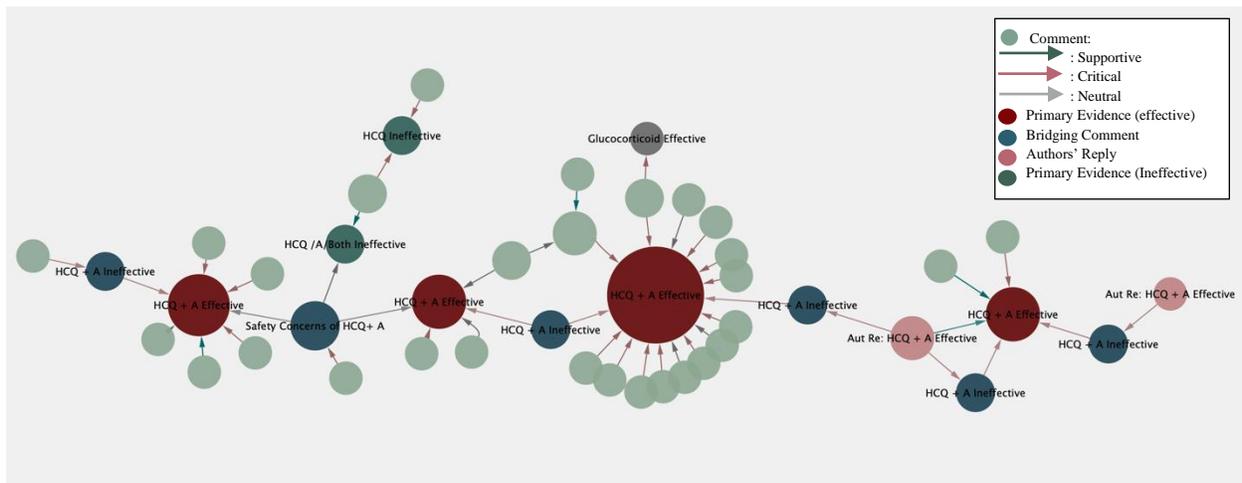

Fig. 7. Treatment efficacy of HCQ + A on COVID-19 was negated

From the figure, all four HCQ + A early research articles concluded that HCQ + A was effective on COVID-19, especially Gautret's team published two articles to demonstrate this view [43-46]. By contrast, significant bridging reviews either attacked the efficacy of this combination on COVID-19 or expressed safety concerns of this off-label usage, except two authors' replies to defend their positions [47-52]. Furthermore, over two-thirds of comments (75.6%) criticized (red arrows) those treatment effectiveness articles, such as case report evidence of inefficacy and concerns of this treatment [47 49 53].

Significantly, when Alizargar commented on two articles claiming the efficacy of HCQ + A, he also commented on Rosenberg et al.'s research who asserted the inefficacy of HCQ/A/Both to integrate conflicting studies and evidence [47 54]. In this way, comment completed evidence appraisal by criticizing problematic evidence and providing related evidence to strengthen own position and thus connected relevant evidence forming ECNs. In the end, an HCQ + A treatment knowledge path was achieved from "efficacy found" to "efficacy negated" due to comments with challenging viewpoints.

The third version of WHO guideline on 17 Dec 2020, firstly recommended against using HCQ on COVID-19, disregarding the severity of disease [33]. This was published posterior to the WHO SOLIDARITY trial on 15 Oct 2020, revealing the inefficacy of HCQ [39]. Rather, the WHO guideline was supported by a "living systematic review and network meta-analysis", which concluded HCQ + A do not seem to be beneficial in treating COVID-19 based on 5 clinical trials, 507 participants, including this intervention, with beneficial treatment evidence level as very uncertainty [55]. It was consistent with the results of our HCQ + A treatment evidence-comment network. In the network, the newest article on June 2020 indicated the uncertain effects of both HCQ solely and HCQ + A. Even though the WHO guideline did not directly refer to the papers (often the most original and preliminary evidence) in our network, this, in turn, proved the hypothesis that ECNs could be used as indirect evidence and early signs of the effect to clue the development of clinical guidelines.

3.4. Corticosteroid

As a treatment candidate of COVID-19, corticosteroid, has been suggested the potent anti-inflammatory effects to mitigate the deleterious effects, such as lung injury and multisystem organ dysfunction, caused by the systemic inflammatory response of severe COVID-19 [56]. In contrast to direct antiviral effect, evidence of corticosteroid more focused on its safety and peripheral effects than virus-related efficacy. Therefore, we analyzed the top two sub-graphs of corticosteroid, as the biggest sub-graph only contains three nodes discussing the efficacy of corticosteroid. Fig. 8 exhibited these two subgraphs of corticosteroid.

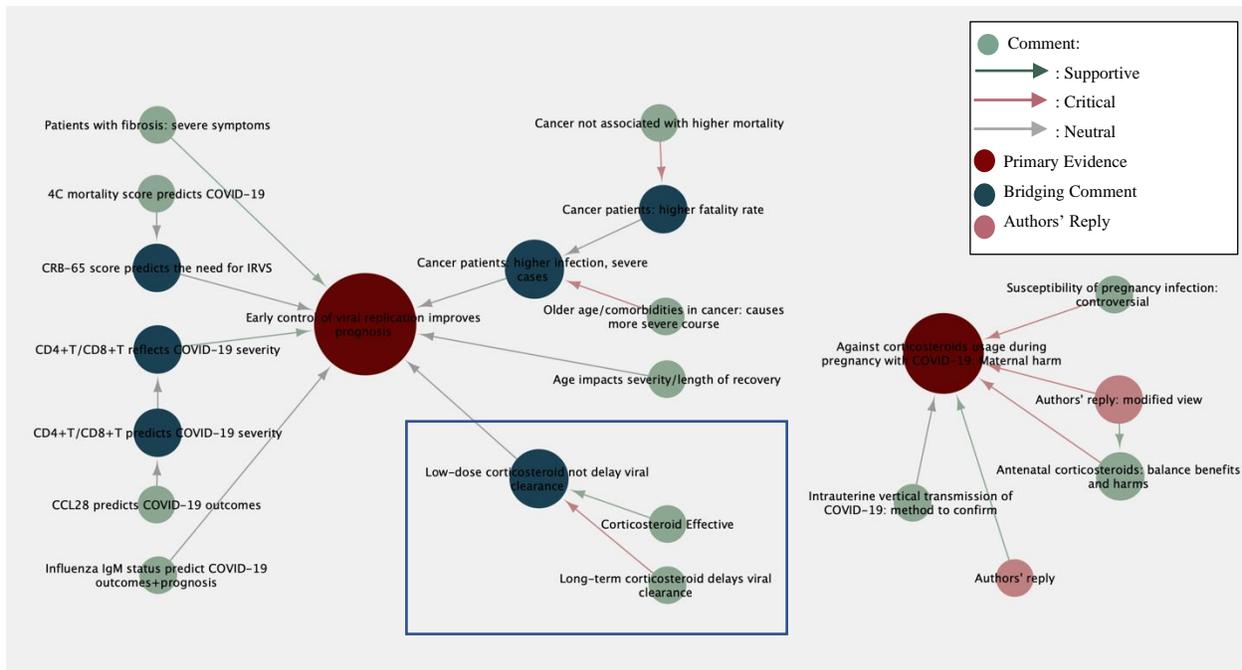

Fig. 8. Treatment efficacy of corticosteroid on COVID-19

Chen et al. elaborated the clinical progression of COVID-19, contending that it was crucial to control viral replication in the early stage and apply host-directed therapy in the later stage to improve the prognosis of COVID-19 [57]. This is the original core research in the biggest weakly connected component involving corticosteroid. Comments of this article mainly talked about the influential factors related to COVID-19. Among them, only one comment explored the treatment of COVID-19 and suggested that "low-dose of corticosteroid therapy may not delay viral clearance", as fast viral clearance is an essential factor of recovery [58]. This suggested corticosteroid as a potential candidate for treating cytokine storm syndrome of patients with severe COVID-19, without the concern of the delay of viral clearance.

However, this commentary received two conflicting comments afterwards. Boglione et al. disclosed their experience with 149 patients; they found that corticosteroid treatment on time has a positive effect on patients before the acute respiratry distress syndrme development, which got a consensus with the research of Chen et al. [59]. By contrast, Jung et al. disagreed with this comment, and they reported a case of a critically ill patient with COVID-19 in Korea detecting viral shedding more than six weeks from her nasopharyngeal swab after receiving low-dose corticosteroid, undermining the effect of corticosteroid [60]. It is hard to conclude the effect of corticosteroid on COVID-19. Thus, we went through the second biggest sub-graph for more evidence-comment information.

The positive effect of corticosteroid on COVID-19 was revealed from the 2$^{nd}$ biggest sub-graph. It began with an article about the antenatal administration of COVID-19 patients, mentioning that against the antenatal corticosteroid therapy for fetal lung maturity during pregnancy of patients with COVID-19, considering deleterious effects corticosteroid on COVID-19 [61]. This article received five comments, and two of them discussed the usage of corticosteroid, including an authors' reply. Liauw et al. firstly refuted the conclusions of maternal harm since the impact of corticosteroid in nonpregnant was unclear; secondly, they argued that the absolute benefits of antenatal corticosteroid vary per week since the baseline risks of neonatal morbidity decreased as gestational age increased [62]. Then, interestingly, the authors replied and modified their original view with the new evidence of the efficacy of corticosteroid, recommended to use corticosteroid during pregnancy with the preterm risk to improve the neonatal outcomes [63 64]. The reversion made the 2nd sub-graph determined the benefits of corticosteroid on COVID-19. Here, as this comment provided new evidence to the original research evidence which reversed the initial assertion, comment-derived evidence created.

Based on the top 2 sub-graphs, the potential of corticosteroid on COVID-19 was detected, challenged, and further confirmed, and this may further evolve as new studies emerge. In addition, this finding corresponded to the first version of WHO guideline, which recommended to use corticosteroid on severe patients and conditionally recommended against using corticosteroid for non-severe patients. Corticosteroid was usually used to fight cytokine release syndrome (CRS), which would develop into severe or critically ill COVID-19 patients.

3.5. Ivermectin

There are seven sub-graphs related to ivermectin. After removing comment nodes not found or in Spanish, the top 2 biggest sub-graphs out of 7 were analyzed in detail, as shown in Fig. 9. Since the nodes' number of the first biggest sub-graph is so small and no clear conclusion could be detected, we displayed the 2nd sub-graph for supplementing the effectiveness analysis of ivermectin. As a result, the comment-driven conclusion from the two ECNs was that "ivermectin effective" was untenable.

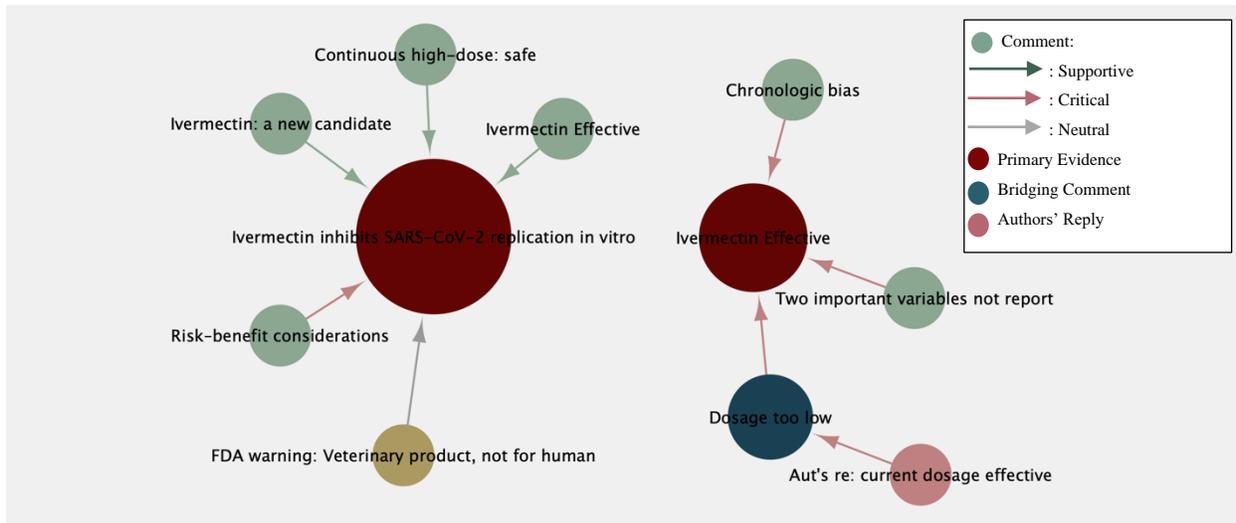

Fig. 9. The treatment efficacy of ivermectin on COVID-19 was untenable

From the subgraph on the left, the seemingly good effectiveness of ivermectin was shown. Caly et al. stated that ivermectin could inhibit the replication of SARS-CoV-2 in vitro and concluded that it needs further research for possible benefits in humans cautiously [65]. This article received five comments in total, among them, three supportive comments, one neutral comment, and one critical comment, which seemed promising from peers' perspectives with only one criticism and three compliments [66] [67] [68] [69] [70].

However, the neutral comment was from the chief editor of *Antiviral Research*, where Caly et al.'s work was published. Bray et al. asserted that this work had been widely spread online and ivermectin being incorrectly interpreted as a treatment drug, thus causing a warning from FDA that "ivermectin in veterinary products should not be used for human therapy" [66]. In sum, the effectiveness of ivermectin cannot be concluded from the ECN on the left. Moreover, the FDA warning emphasized that it was misleading to consider ivermectin as an effective human treatment. The second biggest sub-graph on the right displayed a comment-driven "ivermectin effective: unsolid" assertion. Rajter et al. concluded that ivermectin was associated with lower mortality, especially in patients with severe pulmonary involvement [71]. This article received three critical comments, pointing out a series of methodology errors of the research, including chronology bias, low dosage intervention, not reporting two essential variables that may impact the results (time of symptom onset, patients' health insurance coverage) [72-74]. With all comments negative, the comment-driven conclusion from the network was that the efficacy of ivermectin was untenable.

Based on the top 2 sub-graphs of ivermectin above, the comment-driven conclusion was that ivermectin was not practical for COVID-19. Correspondingly, WHO finally suggested not to use ivermectin for COVID-19 except for clinical trials in the 4th version guideline [19]. This was a super exciting finding. Even though considerable peers were positive, a warning sign could be detected from comments to calm readers down, and this sign could trigger doubt about its validity. After combining to more sub-graphs, a reversed knowledge path of ivermectin would expectedly be detected.

3.6. IL-6 receptor blockers (tocilizumab/sarilumab)

The first biggest sub-graph out of 36 sub-graphs about IL-6 receptor blockers was deeply explored, exhibiting that TCZ (IL-6 receptor blockers) was effective on severe COVID-19 patients in Fig. 10. This was consistent with the latest version of the WHO guideline, which highly recommended using IL-6 receptor blockers in severe patients [19].

Specifically, Xu et al. claimed that tocilizumab (TCZ) was effective to reduce mortality in severe COVID-19 patients, and this article was commented by Yang et al. concerning the safety of the combined use of TCZ and glucocorticoids according to Luo et al.'s research that this combination resulted in disease aggravation [75-77]. Then Xu et al. replied to Yang et al.'s comment and commented on Luo et al.'s research at the same time because Yang et al. mentioned it, explaining Luo et al.'s worse clinical outcomes with the combination may be due to TCZ usage on patients' late stages to reinforce the statement that TCZ was effective [76-78]. This back and forth discussion made the comment and the author's reply as bridging comments connecting two relevant original studies and formed an evidence-comment network for the effectiveness of TCZ.

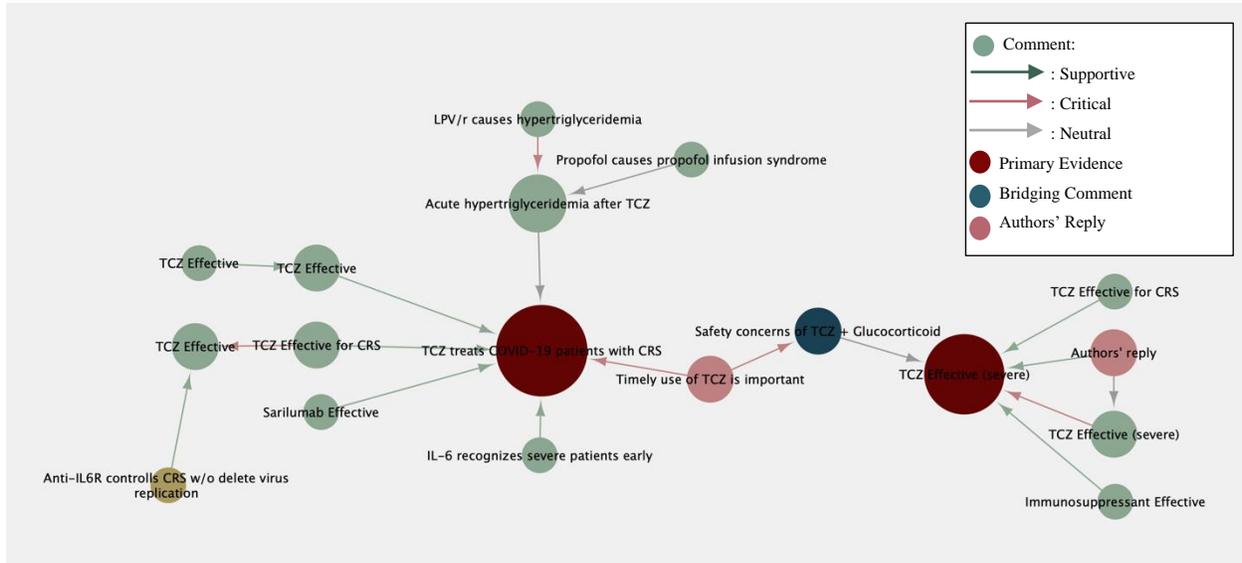

Fig. 10. TCZ (IL-6 receptor blockers) was effective on severe COVID-19

In general, two original studies from Luo et al. and Xu et al. connected by two bridging comments formed the main thread of the graph above [75-78]. The majority of comments (N = 10) of these two articles believed that TCZ/Sarilumab was effective [79-88]. Among them, three comments pointed out that TCZ was effective in cytokine release syndrome (CRS), an acute severe systemic inflammatory response resulting from the exaggerated synthesis of IL-6, which was associated with the SARS-CoV-2 infection [80 81 84]. Also, Buonaguro et al. conceived that the reason why TCZ treated COVID-19 was that it could control CRS instead of decreasing effects of virus replication [80].

Except comments about the effect of TCZ, Morrison et al.'s case report declared that they detected acute hypertriglyceridemia after TCZ usage on COVID-19 patients [89]. However, based on three diagnosed cases, Sharma commented and supplemented that long-term administration of high-dose propofol in intubated COVID-19 patients may cause the incidence of "propofol infusion syndrome", which has the characteristic symptoms of hyperlipidemia and hypertriglyceridemia [90] [91]. Besides, Hassoun et al.'s comment (case report) mentioned the uncertainty of the ideal dose of tocilizumab (IL-6 receptor blockers), confirmed by WHO's latest guideline [19] [84]. This suggested that comments did evidence appraisal to promote the certainty of knowledge by clarifying (resolving) uncertainty as well as alerting (detecting) uncertainty. Specific illustrations are provided in Fig. 11.

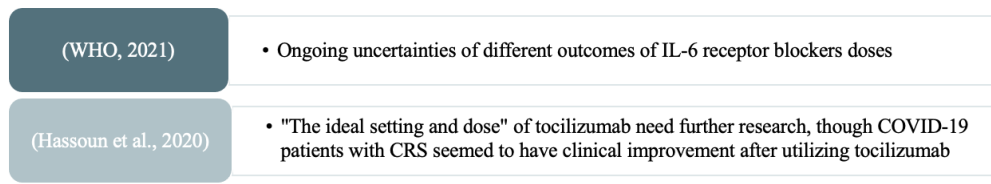

Fig. 11. Uncertainty mentioned both in a comment and then in the 5th version of WHO guideline

4. Comment sentiment and topic analysis

After a detailed elaboration on the biggest two subgraphs, the overall sentimental orientation for all the subgraphs were computed for each drug. The overall sentiment could help further identify the overall treatment propensity of each drug for researchers and clinicians.

4.1. Overall sentiment analysis

When calculating the overall sentiments of each drug, the results were super interesting. Without considering specific topics articles were talking about (i.e., treatment efficacy, inefficacy, or mechanisms of action) and only computing the distribution of comment sentiments for each drug, the results were aligned with recommendations in WHO guidelines.

In specific, WHO guidelines recommended using IL-6 receptor blockers (tocilizumab/sarilumab) and corticosteroid for patients with severe COVID-19. Only IL-6 receptor blockers (48; 25) and corticosteroid (28; 12) received more supportive comments than critical comments, as is shown in Fig. 12. Regarding the rest of drugs, HCQ (62; 94), remdesivir (19; 30), LPV/r (9; 25), and ivermectin (4; 7) respectively received less supportive comments than critical comments. This was consistent with the recommendations for/against using the drug in WHO guidelines.

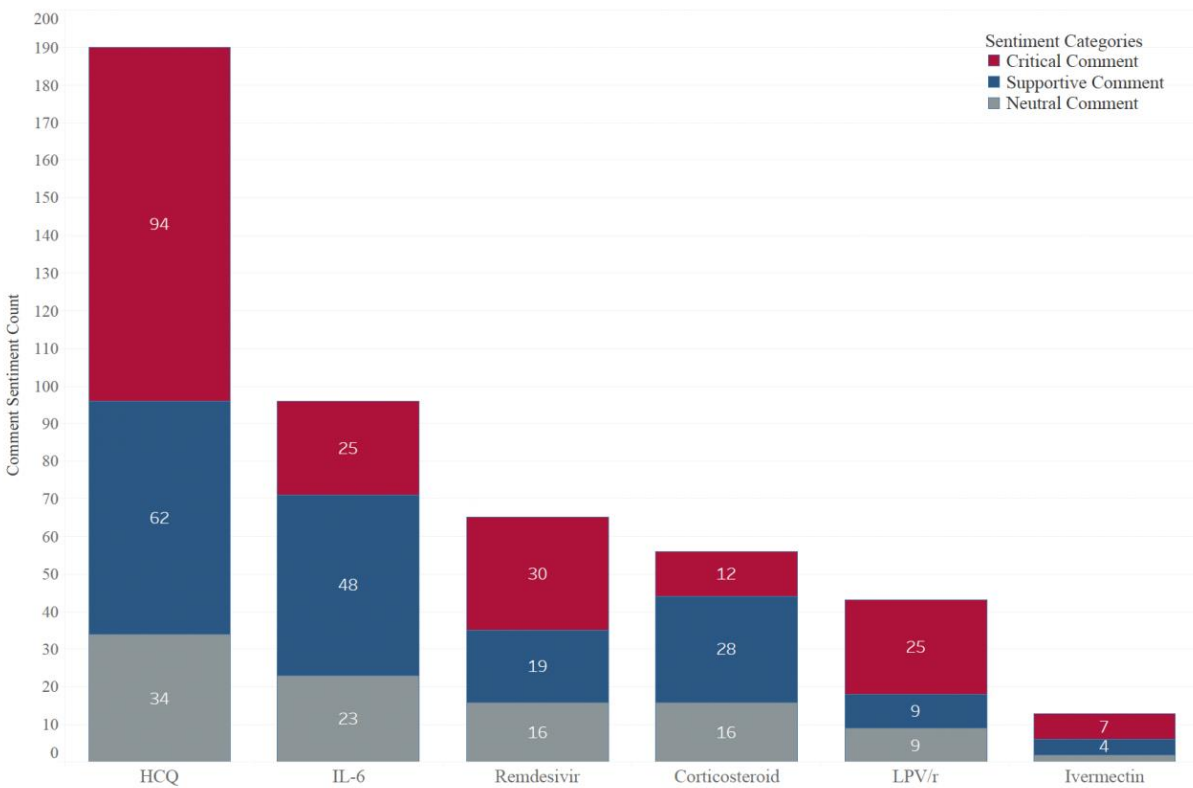

Figure 12. Sentimental orientation of comments for six drugs

4.2. Comment topics analysis

The consistency between comment sentiment and recommendation polarity (for vs against) has been validated above. To probe the sensitivity of comment, we compared the comment topics with the concerns of WHO guidelines and other evidence appraisal systems. We try to uncover whether comment topics were in accord with the core concerns of evidence appraisal criteria in the development of clinical practice guidelines and even beyond.

The quality of evidence, especially the research methodology, is the most crucial determinant in developing guidelines and suggesting recommendations. Nevertheless, having a high quality of evidence does not signify a strong recommendation. Other factors affecting recommendations include clinical applications (significantly the benefit and risk of an intervention), patient values and preference, costs, etc. [92][93]

The distribution of comment topics was plotted in the Fig. 13. The overall distribution showed that the leading comment topic was methodology (54.42%), then followed by clinical themes (31.92%) and other (13.65%). These results aligned with Kastner et al.'s findings [16]. Under each category, the top three sub-categories of methodology topics were analysis (14.04%), intervention (10.38%), and study design (8.46%); in clinical themes topics, these were clinical practice related (11.54%), biological mechanisms (5.00%), and clinical evidence related (3.65%); in other topics, these were just mentioned (8.08%), ethical issues (2.12%), and evidence-based medicine (0.96%).

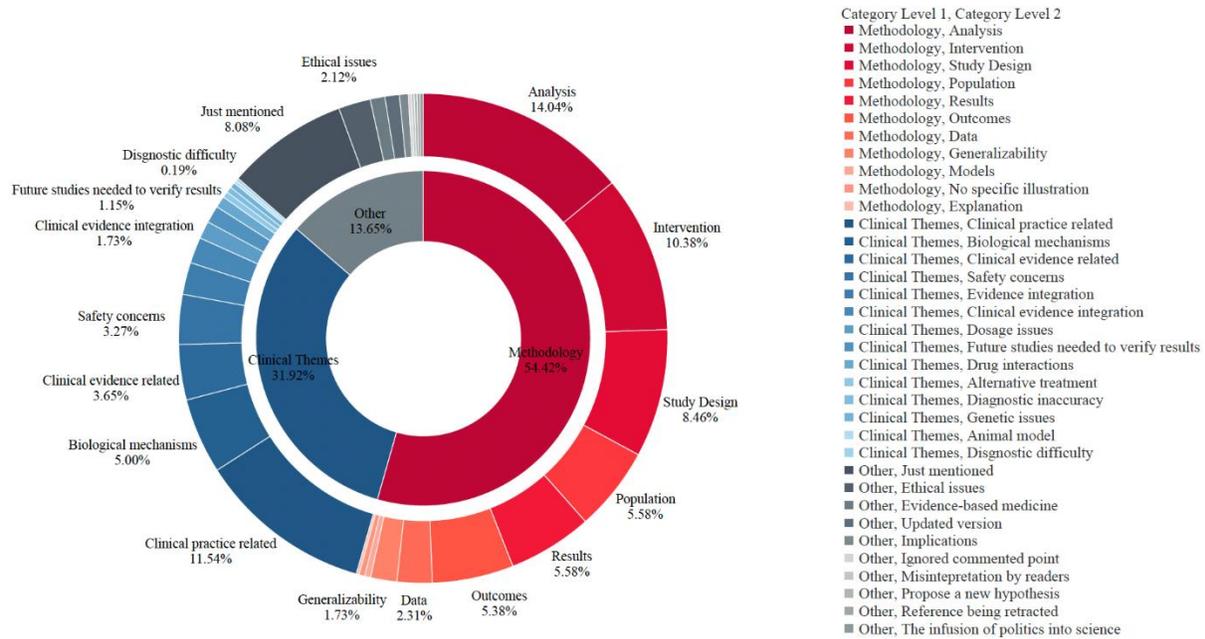

Fig. 13. | The overall distribution of comment topics. The chart includes two layers, and the inner layer was the 1st level of comment topics. The outside layer was the 2nd level topic of each comment under the 1st level.

The above factors affecting recommendations are significantly overlapped with categories in detecting comment topics. Such categories actually form a taxonomy for scientific commentary concerns/points in medical research to help improve evidence appraisal.

Next, to further explore the topics distribution for each drug, we then plotted the sunburst chart of each candidate, respectively, as Fig. 14 below. Interestingly, different from the overall topics' distribution, IL-6 receptor blockers (57.29%, 28.13%) have more comments on clinical themes than methodology. For IL-6 receptor blockers, under clinical themes, the top 1 second-level topic is clinical practice related; this means that many comments talked about the issues of IL-6 receptor blockers in clinical practice. This makes sense since IL-6 receptor blockers were recently detected to be effective for severe patients recently, which corresponds to more clinical experience.

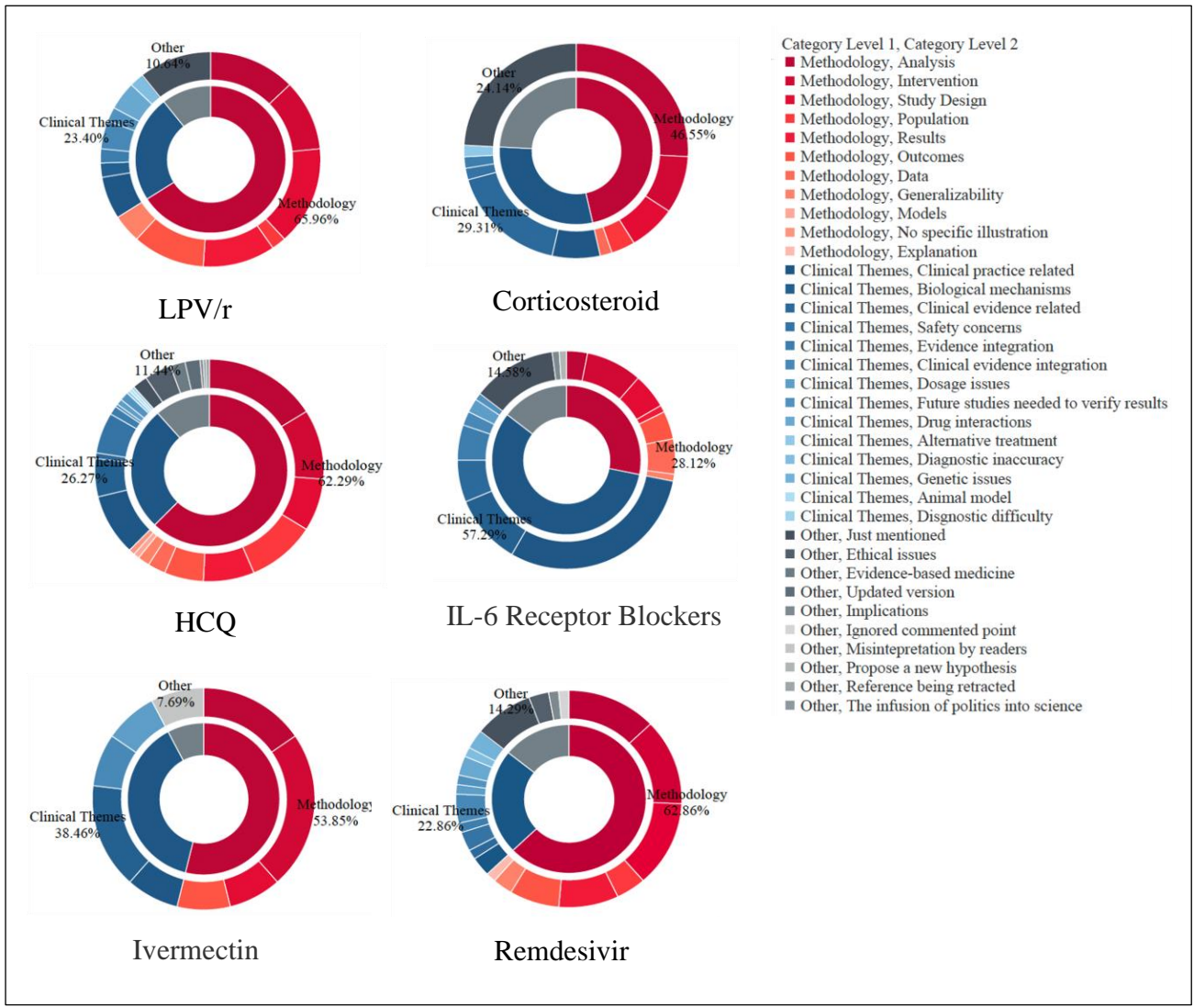

Fig. 14. | The overall distribution of comment topics. The chart includes two layers, and the inner layer was the 1st level of comment topics. Then the outside layer showed the 2nd level of comment topics under the 1st level.

In addition, comments could contribute to various aspects of decision-making beyond science, such as politics, economics, availability, and feasibility. In "other" category of our comment topics, especially in the subcategory of "Just mentioned", where commentors mentioned the research evidence without detailed appraisal but discussing something else related. For example, Self et al. demonstrated that HCQ was ineffective based on their RCT, and Saag commented on this article not to appraise evidence quality but to criticize that the infusion of politics into science resulted in the research craze of HCQ yet no benefits detected [94]. Decision-making is a complex process. Based on different action subjects, different associated factors would be taken into consideration. Comment has the potential to contribute on multiple sides of the process beyond science and beyond evidence quality appraisal.

4.3. Efficiency of comment-driven evidence appraisal

Finally, after analyzing the accuracy and sensitivity of comment-driven evidence appraisal, the most significant part is identifying its timeliness. The comment timespan of each drug was analyzed deeply, especially for each drug, the date of the first published critical comment and the first half of critical comments published compared to the date of WHO guidelines publication. This aims to find to what extent are assertions shaped by critical comments faster than the final released recommendations in guidelines.

We plotted the comment timespan of each kind of sentiment of each drug as Fig. 15. Red sections indicate the timespan of critical comments. For each drug, the first critical comment (red section) emerged earlier than the publication date of WHO guidelines, with an average of 8.8 months. For each drug, the month that half of the critical comments have accumulated at 1) 2020 June & July (corticosteroid, 2.5 months earlier), 2) 2020 May & August (remdesivir, 4.5 months earlier), 3) 2020 July & August (HCQ, 4.5 months earlier), 4) 2020 May (LPV/r, 7 months earlier), 5) 2021

April (ivermectin, 1 month later), and 6) 2020 November (IL-6 receptor blockers, 8 months earlier), respectively. On average there are 4.25 months earlier than WHO guidelines. It indicated that the first half of critical comments emerged, and the critical tendency gradually stabilized and may be applicable for evidence appraisal.

Interestingly, in terms of corticosteroid and IL-6 receptor blockers, all critical comments happened before the guidelines' recommendation, which is intuitive and proved the acceptance of these two drug candidates. Since corticosteroid and IL-6 receptor blockers both fight CRS instead of directly suppressing viral replication, as the safety concerns are solved, the controversy may gradually dissipate. By contrast, HCQ and remdesivir, two controversial candidates kept receiving new critical comments even after the WHO guidelines recommended against using.

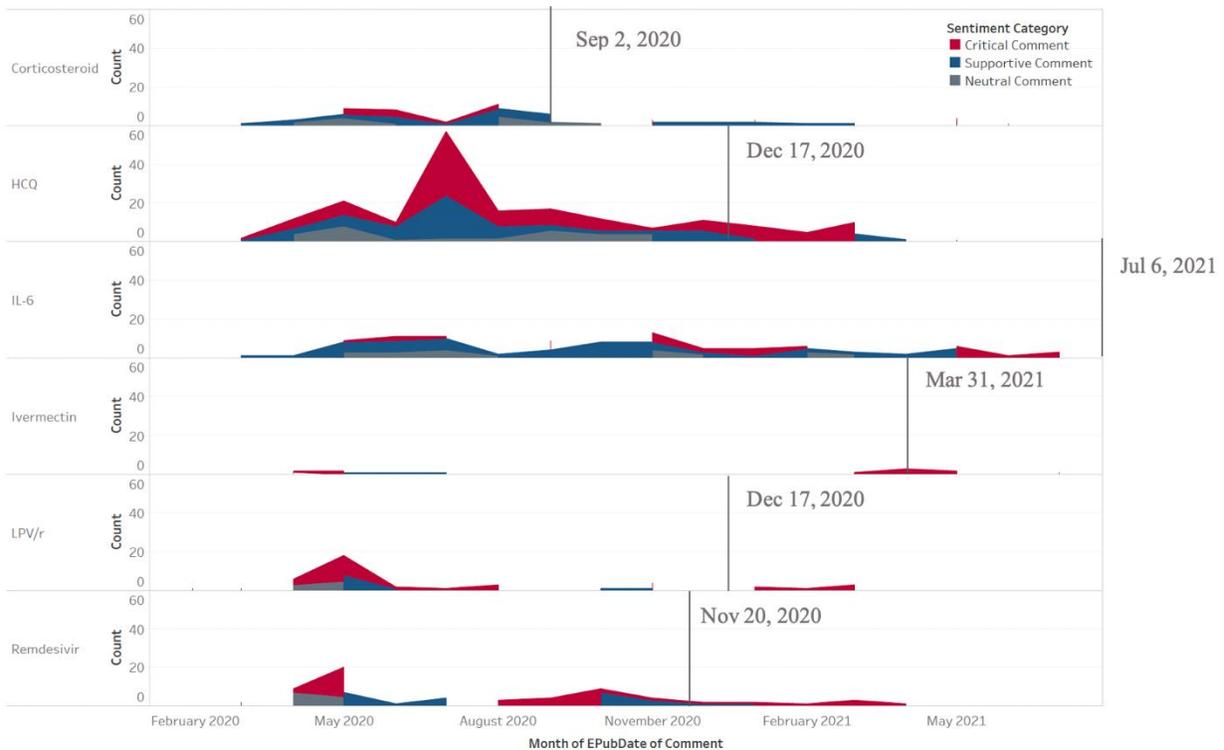

Fig. 15. | Timespan of sentimental orientations for each drug. Vertical reference lines are the publication dates of related recommendations of WHO guidelines.

**Discussion**

The present study aimed to identify how comments appraise clinical evidence and impact the shaping of knowledge. For the accuracy evaluation of comment-driven evidence, except LPV/r, the revealed effectiveness of the biggest subgraphs of the rest 5 drugs were consistent with WHO guidelines, and the overall sentiment for each drug were fully aligned with guidelines. For comment sensitivity analysis, comment topics contained methodology, clinical themes, ethical issues, etc, and matched core concerns when suggesting recommendations in guidelines and even broader. Comment topics went beyond science, and involve politics, society, etc. Finally, for efficiency analysis, half of the critical comments appeared on average 4.25 months earlier than guidelines release. All the above results indicated that comment may serve as an efficient evidence appraisal tool in detecting promising important findings and uncovering unsolid research, and comment could help organize emerging knowledge in emergent situations.

Dr. Horton firstly pointed out that, "Failure to recognize the critical footprint of primary research weakens the validity of guidelines and distorts clinical knowledge" [12]. He stressed the outstanding contribution of research commentaries to shape clinical knowledge, especially criticism, which enlightened our study [12]. After that, Dr. Weng argued the significant role comment played in evidence appraisal [11]. Our study systematically and quantitatively proved their assertions. Comment has the potential as a rapid evidence appraisal tool via providing clues to indicate the importance and validity of evidence for therapeutic evidence-based decision-making, especially in urgent scenarios.

The present study revealed the crucial power of comment-driven evidence assertions by answering three questions, 1) how accurate 2) how sensitive (comprehensive) 3) how fast. Once these three conditions were met, the power of evidence-comment network as an evidence appraisal tool in detecting promising evidence and alerting potential risks in emergent situations was preliminary proved.

1. Accuracy

Regarding the six groups of drug candidates, except LPV/r, the detected effectiveness from the top subgraphs of the rest 5 drugs were consistent with WHO guidelines. Remdesivir, HCQ, and ivermectin were respectively found uncertain or denied efficacy on COVID-19 based on the drug effect relevant evidence-comment paths in their top subgraphs. WHO guidelines recommended against using these drugs as well. For corticosteroid and IL-6 receptor blockers, the top subgraphs revealed the efficacy on severe COVID-19, especially its effect on CSR, and WHO guidelines also recommended using them. Only for LPV/r, comments of the first biggest subgraph considered it a promising candidate, though the second small subgraph of LPV/r thought the effectiveness was uncertain. For evidence validation, guidelines were recommended against using it. As more subgraphs of LPV/ being explored, more information, including reversed evidence, may be disclosed. However, as for the biggest ECNs analysis for each drug, by validating clinical guidelines, the knowledge accuracy of ECNs was 5/6.

In terms of incomplete information of the top biggest subgraphs, the present study further analyzed the overall comment sentiment for each drug. Surprisingly, the overall sentimental orientation fully aligned with the recommendation propensity of WHO guidelines. That is, corticosteroid and IL-6 respectively, have more supportive comments than critical comments, and finally, guidelines recommended using them on severe COVID-19 patients. Remdesivir, HCQ, ivermectin, and LPV/r respectively have more critical comments than supportive comments, and correspondingly guidelines recommended against using them. Even though, based on the LPV/r effect relevant research and comments from its top subgraphs, this combination was considered the potential treatment. The accuracy of comment-driven evidence was proved initially from the above analysis.

Inspired by these findings, we proposed that comment sentiment analysis could be used as a simple, straightforward tool of heuristic decision making, which means making decisions with limited information in urgent situations. In the case of drug effects based on ECNs, locating evidence-comment pairs related to a specific drug, without considering whether the scientific evidence was directly about the effectiveness or not, whether the effectiveness was positive or negative, or whether the comment was directly relevant to the topic or not, just rely on the overall comment sentiment towards the drug and one can conclude. This would be super helpful when only knowledge of uncertainty existed for decision-making. It was a boldly preliminary proposal regarding ECN as an evidence appraisal tool, a more detailed design of the tool would be elaborated below.

2. Sensitivity

Compared to the overall evidence accuracy in decision-making, the sensitivity of comment is significant at the same time, which guarantees the sustainability and promotability of accurate conclusions. This study analyzed the comment topics; topics ranged from methodology, clinical themes to others. Atkins et al. pointed out the criteria to appraise a grading system, and the first standard was "to what extent is the approach applicable to different types of questions (effectiveness, harm, diagnosis, and prognosis)" [95]. Comment fully met this standard since it covered all these points and even beyond this range. Beyond science, comment may involve many aspects of decision-making, such as politics and economics.

In methodology, topics covered the overall research process, containing study design, population, data, intervention, models, outcomes, results, analysis, discussion, and generalizability. In clinical themes, topics covered biology (i.e., biological mechanisms, genetic issues), diagnosis (i.e., diagnostic difficulty, diagnostic inaccuracy), treatment & drug (i.e., alternative treatment, dosage issues, drug interactions, safety concerns), medical evidence (i.e., clinical practice related, evidence integration), other clinical issues (i.e., animal model). In other, topics consist of ethical issues, enlightenment (i.e., propose a new hypothesis), critical thinking (i.e., evidence-based medicine), and other issues (i.e., just mentioned). The scope of comment topics is broader than the four points Atkins et al. mentioned [95].

Moreover, besides the variety of comment topics, the points that commentaries focused on were in accord with the concerns of current grade systems (i.e., GRADE). As an overall clinical evidence appraisal system, GRADE evaluated evidence quality firstly based on study design, RCT/observational research [93]. Study design (i.e., RCT or not), as the 4th common comment topic in our study, was consistent with the concerns' priority of GRADE. In addition, effectiveness, as GRADE's first level up evidence, is the 2$^{nd}$ common topic in our analysis. All these matched points further proved the consistency between comment topics and concerns of grading guidelines.

3. Efficiency

Accuracy and sensitivity are the prerequisites to enable comment-driven evidence appraisal to aid clinical policymaking. However, what makes it the most competitive is the timeliness. First, evidence being commented on means the significance or controversy of the research article. Screening research evidence via "has comment" linkage in PubMed has already filtered out those side articles and induced the appraisal scope. Second, the present study found that the first critical comment was detected 8 months on average before the guideline release. Besides, on average 4.25 months earlier than guidelines, half of the critical comments had already accumulated, which indicated the critical comments is an early sign for the recommendation in guideline.

The biggest issue for clinical guidelines is the slow-developing process. If an approach from the perspective of informatics could help fast detect critical signs or assertions, it would provide hinds for clinical guidelines development. Furthermore, the timeliness made the comment-driven method powerful for evidence appraisal when fast decisions are needed in urgent situations.

4. A quantitative framework by leveraging comment for evidence appraisal

Scientific commentaries are an important way of formal academic community, but these data remain underutilized. Based on the above analysis of evidence-comment networks, comment-driven evidence assertions were able to do evidence appraisal in an accurate, precise, and fast way.

How to take advantage of this tool in a simple but powerful way becomes a worthy discussing issue. Here, according to our research findings and the criteria for a sensitive grading approach of Atkins et al., we proposed to build a grading system based on the comment topics and sentimental orientations of ECNs [95].

This grading tool could appraise evidence from the perspective of comments and help identify those solid new findings or the limitations for the existing evidence which is not pointed out by authors themselves. We propose a comment-driven evidence grading scale from the informatics approach displayed in table 2.

Table 2. A comment-driven evidence grading scale from the informatics approach

| Comment topics | Comment sentiment | Evidence Certainty Levels |
| --- | --- | --- |
| Study Design (2) | | High certainty: [6, 10] |
| Population/Data (2) | Supportive: + 2 | Moderate-high certainty: [2, 6] |
| Analysis (2) | Critical: - 2 | Moderate certainty: [-2, 2] |
| Clinical Practice Related (2) | Neutral: 0 | Moderate-low certainty: [-6, -2] |
| Adverse Effect (2) | | Low certainty: [-6, -10] |

For a given article, each comment was analyzed to identify its comment topics and comment sentiment. All comment topics would then be matched to the scale's comment categories. Once matched to a topic of the scale, the average score of all comments under each topic was the topic score. This avoided the problem of incomprehensive comment topics. In other words, an article received many comments, but they all concentrated on one topic. The final overall score is the sum of each topic score.

Therefore, to take advantage of this tool efficiently, we suggested that medical journals label comment sentiment and topics once accepted comments for publishing. This could help create structured data for calculating comment-driven evidence assertions. Dr. Chunhua Weng has expressed a similar view on *the Lancet* by using controlled vocabulary, such as "lacks equipoise", to re-present journal articles (comments) [11]. This could be used to build a knowledge base to connect clinical research and their comments in a structured format to leverage the potential of scientific commentaries in evidence appraisal.

5. Limitations and future research

Despite multiple remarkable new findings, this article has some limitations as well. Firstly, this study only considers PubMed publications. In addition, whether comments are made without any conflicting interests is not considered. If such a comment on behalf of a particular stakeholder exists, its effectiveness in evidence appraisal should be weak because it loses objectivity.

For the next step, based on this current study, we will extract all randomized clinical trials (RCTs) and their comments from PubMed for evidence-comment network analysis. A dynamic clinical knowledge platform built from evidence-comment networks will be established with the analysis results. This could serve as a clinical knowledge base from which healthcare providers and researchers can efficiently consolidate and expand field knowledge through given clinical ECNs based knowledge paths. Furthermore, this platform could serve as a crucial alert and supplement for developing clinical practice guidelines.

**Conclusions**

This study intended to explore whether and how research commentaries could be used as an evidence appraisal tool cluing evidence-based decision making, especially fast decisions. We extracted evidence-comment data of 6 groups' COVID-19 drug candidates on PubMed as a case study, including 168 evidence publications (146 primary research articles) and 376 comments, forming 427 evidence-comment pairs. It turned out that commentary played a pivotal role in evidence appraisal and can contribute to developing clinical guidelines and decision-making in urgent situations like the COVID-19 pandemic.

Specifically, the study conducted evidence-comment network analysis, sentiment analysis, comment topics analysis, and efficiency analysis. First, for ECNs analysis, out of 6 drug candidates, comment-driven evidence assertions of leading subgraphs of 5 drugs were consistent with WHO clinical therapeutics guidelines recommendations. The sentiment analysis further revealed that the overall comment sentiment of 6 drugs respectively corresponds with WHO clinical guidelines. This proved the accuracy of comment-driven evidence. Second, comment topics covered all significant points of evidence appraisal, from methodology, clinical themes, to other topics (i.e., ethical issues). What is more, the core topics that comment concentrated on, such as study design, were matched up with the concerns of guidelines and other evidence appraisals and even went further to politics, economics, etc. These findings revealed the comprehensiveness and precision of comment topics. Finally, the first half of critical comments emerged 4.25 months earlier on average than the release date of WHO guidelines, which made comment-driven assertions a time-competitive appraisal tool.

Based on the accuracy, sensitivity, and efficiency performance, comment has the potential as a rapid evidence appraisal tool via providing clues to indicate the importance and validity of evidence for therapeutic decisions. Comment has a selection effect by appraising the benefits, limitations, and other clinical practice issues of concern for existing evidence. Although negative comments are critical, they tend to provide a more detailed understanding of the results by introducing new views or evidence, and further shaped the knowledge assertions, rather than simply denying the key findings contained in the initial evidence. Comments also have the potential to inform decision-makers from various perspectives beyond science, such as economics, politics, and ethical issues, which are crucial aspects in policy making.

## Acknowledgement

Shuang Wang (SW) conducted analysis and wrote the manuscript. Jian Du (JD) proposed the conceptual framework and revised the manuscript. We thank Qianying Guo (QYG) for annotating the comment topics and sentimental orientations, and Daoxin Yin (DXY) for providing constructive discussions. We thank any anonymous reviewers. This work was funded by the National Natural Science Foundation of China (71603280, 72074006), Peking University Health Science Center and the Young Elite Scientists Sponsorship Program by China Association for Science and Technology (2017QNRC001).